\documentclass[12pt]{spieman}  
\usepackage{amsmath,amsfonts,amssymb}
\usepackage{graphicx}
\usepackage{setspace}
\usepackage{tocloft}

\title{Video-rate dual-modal forward-viewing photoacoustic and fluorescence endo-microscopy through a multimode fibre}

\author[a]{Tianrui Zhao}
\author[a]{Michelle T. Ma}
\author[a]{Sebastien Ourselin}
\author[a]{Tom Vercauteren}
\author[a,*]{Wenfeng Xia}
\affil[a]{School of Biomedical Engineering and Imaging Sciences, King’s College London, $4^{th}$ Floor, Lambeth Wing St Thomas’ Hospital, London SE1 7EH, United Kingdom}

\cftpagenumbersoff{figure}
\cftpagenumbersoff{table} 
\begin{document} 
\maketitle

\begin{abstract}
Multimode fibres are becoming increasingly attractive in optical endoscopy as they promise to enable unparalleled miniaturisation, spatial resolution and cost as compared to conventional fibre bundle-based counterpart. However, achieving high-speed imaging through a multimode fibre (MMF) based on wavefront shaping has been challenging due to the use of liquid crystal spatial light modulators with low frame rates. In this work, we report the development of a video-rate dual-modal forward-viewing photoacoustic (PA) and fluorescence endo-microscopy probe based on a MMF and a high-speed digital micromirror device (DMD). Light transmission characteristics through the fibre were characterised with a real-valued intensity transmission matrix algorithm, and subsequently, optimal binary patterns were calculated to focus light through the fibre with wavefront shaping. Raster-scanning of a tightly focused beam (1.5 $\mu$m diameter) at the distal end of the fibre was performed for imaging. With the DMD running at 10 kHz, the PA imaging speed and spatial resolution of were controlled by varying the scanning step size, ranging from 1 to 25 frames per second (fps) and from 1.7 to 3 $\mu$m, respectively, over a field-of-view of 50 $\mu{m}$ $\times$ 50 $\mu{m}$. High-resolution PA images of carbon fibres, and mouse red blood cells were acquired through a MMF with high image fidelity at unprecedented speed with MMF-based PA endoscope. The capability of dual-modal PA and fluorescence imaging was demonstrated by imaging phantoms comparing carbon fibres and fluorescent microspheres. We anticipate that with further miniaturisation of the ultrasound detector, this probe could be integrated into a medical needle to guide minimally invasive procedures in several clinical contexts including tumour biopsy and nerve blocks.

\end{abstract}


{\noindent \footnotesize\textbf{*}Wenfeng Xia,  \linkable{wenfeng.xia@kcl.ac.uk} }

\begin{spacing}{2}   

\section{Introduction}
\label{sect:intro}  



Optical endoscopy is widely used for visualising internal organs such as the gastrointestinal tract, the airways and the uterus \cite{wang2004optical}. However, conventional optical endoscopy only provides morphological information of superficial tissue, and thus biopsies are still required to achieve definitive histopathological diagnosis. Owing to the rapid development of optical imaging techniques, several optical endoscopy modalities have emerged as promising optical alternatives to biopsy, including endoscopic optical coherence tomography (EOCT) \cite{gora2017endoscopic,tsai2017optical}, fluorescence endoscopy (FM) \cite{mavadia2012all,flusberg2005fiber}, confocal endomicroscopy (CEM) \cite{delaney1994fiber,gmitro1993confocal} and PA endoscopy (PAE) \cite{zhao2019minimally,zhao2018optical}. EOCT provides real-time 3D images of tissues with microscopic scale (cellular) structural information derived from optical scattering of biological tissue, however, measuring functional information such as blood oxygen saturation with EOCT is challenging. FM and CEM provide anatomical and pathological information of tissues and cells, but it is difficult to perform in-depth imaging as EOCT. PAE provides 3D cellular-level imaging of internal organs in real-time, with contrast that derives from optical absorption by intrinsic tissue chromophores such as haemoglobin, lipids, and extrinsic contrast agents \cite{zhao2019minimally,beard2011biomedical}. Utilising excitation light with multiple wavelengths, multispectral PAE images can be acquired to obtain spatial distributions of absorbing chromophores and contrast agents \cite{ntziachristos2010molecular,xu2006photoacoustic}. It is therefore ideally suited to visualise changes in vascular morphology and blood oxygenation and metabolism that are known to be associated with tumour development \cite{ntziachristos2010molecular,xu2006photoacoustic,mallidi2011photoacoustic}. 

Most of the current PAE probes are based on side-reviewing configurations, with which volumetric images of tissue are acquired by pull-back of the rotating probes. These probes are useful for visualising hollow structures such as intravascular \cite{jansen2011intravascular,wu2017real} and gastrointestinal tract \cite{yang2012simultaneous} imaging where circumferential scanning is required. In the past few years, forward-viewing PAE probes have attracted significant research interest as they could be more convenient for use in various clinical applications such as optical biopsy and tumour margin assessment compared to side-viewing probes \cite{hajireza2011label,hajireza2013label,ansari2018all}. Coherent fibre bundles are commonly used in PAE probes in both microscopy and tomography modes. With PA microscopy mode, a MEMS or a galvanometer mirror system was used to scan a focused laser beam at the proximal tip of the fibre bundle into individual cores, whilst the generated ultrasound (US) A-scans were received by a single-element ultrasound (US) transducer \cite{hajireza2011label}. The imaging capability was demonstrated with mouse ears \textit{in vivo}, however, the lateral resolution of the optical-resolution PAE probe was 7-8 $\mu$m, limited by the large separation between individual cores of the fibre bundle \cite{hajireza2011label}. PA tomography via a coherent fibre bundle was extensively studied by the group of Beard with dichromatic Fabry-Perot sensors for US detection and the system development has undergone several iterations \cite{ansari2018all,ansari2020miniature}. In 2018, Ansari et al. developed a PAE system by coating a transparent Fabry-Perot sensor at the tip of a 3.2 mm-diameter rigid fibre bundle and raster-scanning an interrogation laser beam through the bundle for US detection \cite{ansari2018all}. In 2020, the same group developed a flexible PAE probe with a diameter of 7.4 mm by interrogating a Fabry-Perot sensor through a flexible fibre bundle with a miniature optical relay system \cite{ansari2020miniature}. With these probes, high-quality 3D vasculature images of duck embryo and human placenta \textit{ex vivo} were obtained. However, the diameters of these probes were still too large to be integrated into a biopsy needle and the imaging speed was too slow for clinical operation (100s to 25 min per image).

Recently, multimode optical fibres (MMFs) have shown great promise for use in forward-reviewing optical endoscopy owing to the rapid advances of optical wavefront shaping technology \cite{turtaev2018high,choi2012scanner,loterie2015digital,vasquez2018subcellular,caravaca2017single,ohayon2018minimally,papadopoulos2013high,morales2015two,papadopoulos2013optical,gusachenko2017raman}. Compared to coherent fibre bundles, MMFs are much more cost effective. Further, as the sizes of optical foci through a MMF are limited by optical diffraction, the effective pixel density and achievable spatial resolution with a MMF can be 1-2 orders of magnitudes higher \cite{andresen2016ultrathin}, and as such high-resolution MMF-based probes can be readily miniaturised for integration into medical needles. With MMF-based optical endoscopy, a focused light beam is raster-scanned at the distal end of a MMF via wavefront shaping with the knowledge of the light transmission characteristics of the MMF, which are usually achieved with transmission matrix-based approaches \cite{vasquez2018subcellular,vcivzmar2011shaping} or digital optical phase conjugate \cite{papadopoulos2013high,papadopoulos2013optical} approaches using a liquid-crystal spatial light modulator (LC-SLM). In the past decade, several groups have developed ultrathin endoscopes based on MMFs and wavefront shaping for a wide range of biomedical imaging modalities including wide-field microscopy \cite{choi2012scanner}, confocal \cite{loterie2015digital}, fluorescence \cite{turtaev2018high,caravaca2017single}, two-photon \cite{morales2015two}, Raman \cite{gusachenko2017raman}, PA imaging \cite{papadopoulos2013optical} and multi-modal imaging probes \cite{mezil2020single}. However, due to the slow rates of LC-SLM (~100 Hz) and data acquisition, the fastest MMF-based PAE system reported in literature required 30 s for the acquisition of a single PA image frame comprising 1800 pixels \cite{mezil2020single} and hence hinders its clinical translation. In contrast to the LC-SLM, a digital micromirror device (DMD) that consisted of a large array of micromirrors, offers binary amplitude modulations by switching ‘ON’ or ‘OFF’ micromirrors at a high frame rate of 23 kHz. High-speed fluorescence imaging of a mouse brain \textit{in vivo} through a MMF using DMDs was demonstrated with a frame rate of 3.5 fps for 7000-pixel images \cite{turtaev2018high} and it can be further improved to 7-15 fps with sub-sampling \cite{ohayon2018minimally}. 


In this work, we developed, for the first time to our knowledge, a high-speed, MMF-based PA and fluorescence endomicroscopy system using a DMD. The performance of two types of MMFs, a step-index (STIN) and a gradient-index (GRIN) fibre, were compared and the GRIN fibre was preferred due to its more uniform intensity distribution of the optical foci. As the scanning step size can be easily varied using a DMD, the lateral resolution and the imaging speed of the developed PAE system is scalable. With a carbon fibre phantom, the measured spatial resolution of the PA images ranged from 1.7 to 3 $\mu$m with the step size varied from 0.5 to 2.5 $\mu$m, and the corresponding imaging speed with a field-of-view of 50 $\mu{m}$ $\times$ 50 $\mu{m}$ varied from 1 to 25 frames per second (fps), respectively. Further, biconcave structures of mouse red blood cells (RBCs) were visualised with PA imaging. Finally, the capability of dual-modal forward-reviewing PA and fluorescence imaging of the endo-microscopy probe was demonstrated with a phantom comprising a carbon fibre and fluorescence microspheres.

\section{Materials and methods}
\subsection{Imaging system}
A schematic diagram of the imaging system is shown in Fig. 1. A pulsed laser (532 nm, 2 ns, SPOT-10-200-532, Elforlight, Daventry, United Kingdom) was used as the light source for both PA and fluorescence imaging. A DMD with 768×1080 pixels (DLP7000, Texas Instruments, Texas, USA) was used to project binary patterns onto the proximal end of a MMF via a tube lens (AC254-050-A-ML, Thorlabs, New Jersey, USA), a circular polariser (CP1L532, Thorlabs, New Jersey, USA) and an objective (20$\times$, 0.4 NA, RMS20X, Thorlabs, New Jersey, USA). Two types of multimode fibres including a step-index (STIN, Ø105 µm, 0.22 NA) and a gradient-index (GRIN,  Ø100 µm, 0.29 NA) ones with the same length of 20 cm were employed. A sub-region of the DMD covering 128 $\times$ 128 micromirrors was used for light modulation with a frame rate of 10 kHz. Prior to image acquisition, a fibre characterisation step was performed by a fibre characterisation unit that comprised a CMOS camera (C11440-22CU01, Hamamatsu Photonics, Shizuoka, Japan) for capturing the output speckle patterns after magnification by an objective (20$\times$, 0.4 NA, RMS20X, Thorlabs, New Jersey, USA) and a tube lens (AC254-0100-A-ML, Thorlabs, New Jersey, USA). The focal plane of the camera was set to approximately 100 $\mu$m away from the distal end of the fibre. 


\subsection{Multimode fibre characterisation}
\label{sect:title}
A real-valued intensity transmission matrix (RVITM)-based method was used for fibre characterisation with a DMD. The method has been detailed in previous studies \cite{Zhao:20,zhao2021high}, here briefly, a Hadamard matrix H $\in$ (-1, +1) with dimensions of N $\times$ N was generated in MATLAB, then two binary matrices \(H_1 = (H+1)/2\) and \(H_2 = (-H+1)/2\) were constructed based on H and combined as a binary Hadamard matrix $[H_1,H_2]$ $\in$ (0, 1). All the columns of the matrix $[H_1,H_2]$ were then converted into square patterns, which were sequentially displayed on the DMD whilst the intensity distributions of the output speckles at the distal fibre tip were captured. It was found that intensity changes from all input modes (DMD micromirrors) to output modes (camera pixels) could be modelled as \cite{Zhao:20,zhao2021high}:

\begin{equation}
    \begin{bmatrix}
    I^1_1 & \cdots & I^{2N}_1 \\
    \vdots & \ddots & \vdots \\
    I^1_m & \cdots & I^{2N}_m\\
    \end{bmatrix} = RVITM \bullet {[H_1,H_2]},
\end{equation}
where $I^k_m$ is the intensity at the $m^{th}$ output mode when the $k^{th}$ binary Hadamard pattern is displayed as input, $N$ is the total number of input modes. Then, by taking advantage of the properties of the Hadamard matrix, we have $[H, -H]^T = [H, -H]^{-1}$, and the value of RVITM can be calculated via:

\begin{equation}
    RVITM = \begin{bmatrix}
    2I^1_1-I^1_1 & \cdots & 2I^{2N}_1-I^1_1 \\
    \vdots & \ddots & \vdots \\
    2I^1_m-I^1_m & \cdots & 2I^{2N}_m-I^1_m\\
    \end{bmatrix} \bullet {[H,-H]^T},
\end{equation}

As a result, a positive $rvit_{mn}$ (the intensity transmission constant linking the $n^{th}$ input mode and the $m^{th}$ output mode) indicates the improvement of intensity at the output mode when a micromirror is switched ‘ON’. Furthermore, $rvit_{mn}$ can be further expressed as: 
\begin{equation}
    rvit_{mn} = A_{mn}A_Rcos(\theta_{mn} - \phi_R)
\end{equation}
where $A_{mn}$ and $\theta_{mn}$ are the amplitude and phase at the $m^{th}$ output mode when only the $n^{th}$ micromirror is switched ‘ON’, $A_R$ and $\phi_R$ are the amplitude and phase of the output light field when all micromirrors are switched ‘ON’, respectively \cite{zhao2021high}. So, switching ‘ON’ micromirrors with positive $rvit_{mn}$ values leads to constructive interference because the phases of output fields are in the range of $[-\pi/2,\pi/2]$. Furthermore, we ranked all the micromirrors descendingly with their $rvit_{mn}$ values and switched ‘ON’ the top 30$\%$ micromirrors in the optimal patterns to achieve the highest peak-to-background ratio for focusing at desired spatial locations at the distal end of the MMF, as we demonstrated in Ref. \cite{zhao2021high}.

\subsection{Image acquisition}
After fibre characterisation, the fibre characterisation unit at the distal fibre tip was replaced by a PA imaging unit (Fig.1), which comprised a single-element piezoelectric US transducer (V358, central frequency: 50 MHz; diameter: 0.25 inches, Olympus, Japan) and a cover slip as a sample holder that were integrated into a custom imaging chamber. An acoustic lens (LC4210, f = -25 mm, Thorlabs, New Jersey, USA) was attached to the active surface of the US transducer for focusing at the optical focal plane of the MMF. Both the US transducer and the distal end of the MMF were affixed to two 3D translation stages to facilitate precise alignment between the focus of the US transducer, the focal plane of the MMF and the imaging targets. The diameter of the acoustic focal area at the focal plane was estimated to be $\sim$ 50 $\mu$m. Imaging targets, including carbon fibre phantoms and mouse RBCs, were placed on the sample hold located at the optical focal plane. The imaging chamber was filled with deionised water for acoustic coupling.


By displaying optimal DMD patterns, the excitation light was focused through the MMF and raster-scanned over the imaging targets. For PA imaging, the generated US waves were received by the US transducer, which were then amplified (SPA.1411, Sprectrum Instrumentation, Grosshansdorf, Germany), digitised by a data acquisition card (M4i.4420, Sprectrum Instrumentation , Grosshansdorf, Germany) and transferred to a personal computer (Intel i7, 3.2 GHz) for processing and display. For fluorescence imaging, the excited fluorescent light was collected by the MMF and captured by a photodetector (400-1000 nm, 10 MHz, APD410A/M, Thorlabs, New Jersey, USA) after reflected by a beam splitter (BS013, Thorlabs, New Jersey, USA) and spectrally filtered by a dichroic mirror (FEL0550, Thorlabs, New Jersey, USA). The received fluorescence signals were then sent to a second channel of the data acquisition card and transferred to the personal computer for processing and display. Synchronization of the imaging system was controlled by a functional generator (33600A, Keysight, Santa Rosa, California) and a custom MATLAB program.



\begin{figure}
\begin{center}
\begin{tabular}{c}
\includegraphics[width=16cm]{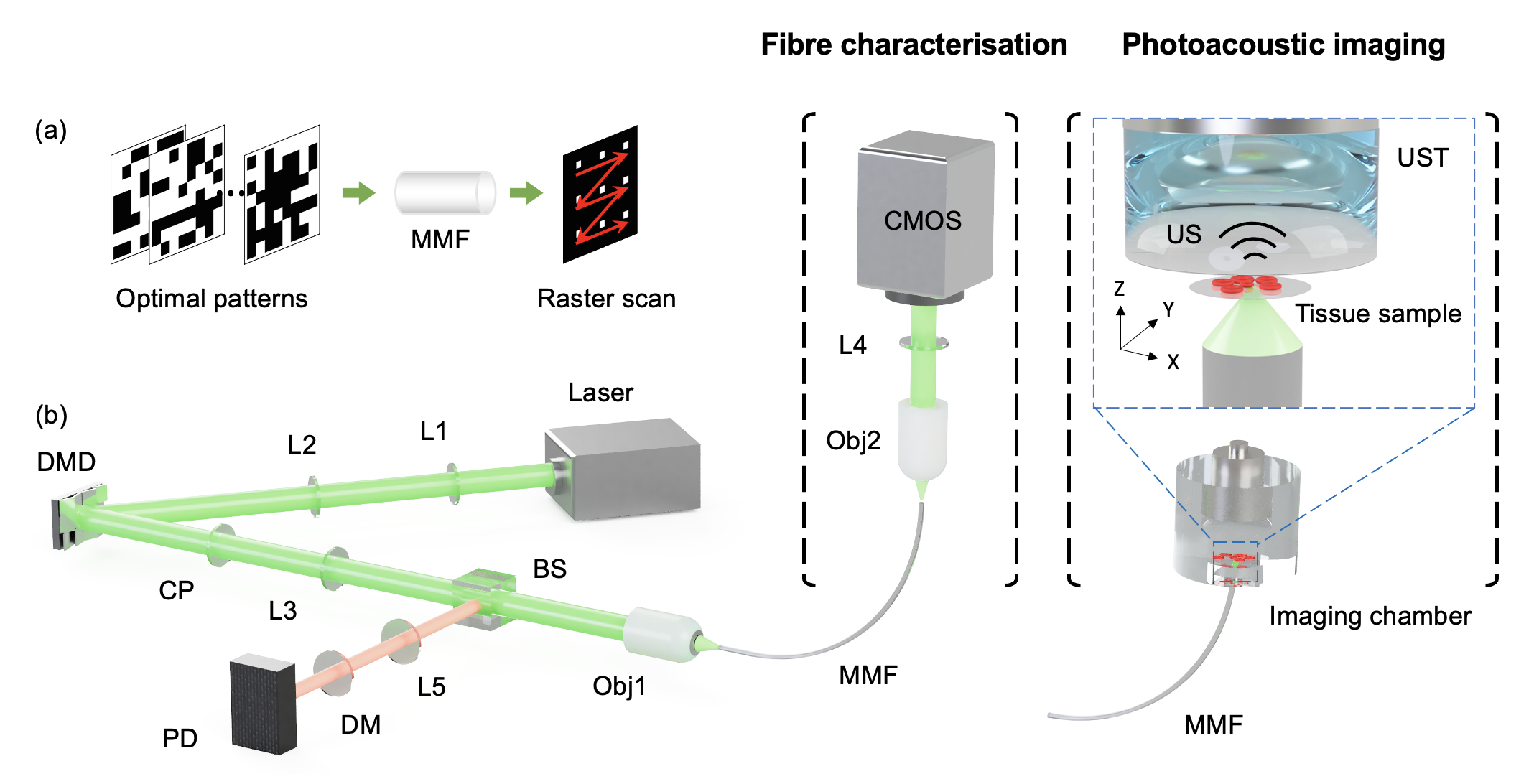}
\end{tabular}
\end{center}
\caption 
{ \label{fig:example}
Schematic diagrams of the principle and configuration of the video-rate dual-mode forward-viewing photoacoustic and fluorescence endo-microscopy system. (a) The principle of raster-scanning of a focused light beam through a multimode fibre (MMF) for endo-microscopy imaging by sequentially displaying optimal DMD patterns via optical wavefront shaping. (b) Configuration of the imaging system; L2, tube lens (f = 50 mm); L3-5, tube lenses (f =  100 mm);  DMD: digital micromirror device; CP, circular polariser; DM, dichroic mirror (longpass Filter, cut-on wavelength: 550 nm); CMOS: complementary metal-oxide-semiconductor camera; US, ultrasound; PD: Photodetector; UST, ultrasound transducer; Obj1-2: Objective lenses.} 
\end{figure} 

\subsection{Experiments}
Several experiments were performed to evaluate the performance of the imaging system. First, the focusing performance with the RVITM-based method was evaluated for two MMFs: a STIN fibre with a core diameter of 105 $\mu$m, and a NA of 0.22, and a GRIN fibre with a core diameter of 100 $\mu$m, and a NA of 0.29 NA. Three performance metrics were compared: the enhancement factor (EF), the size of the focus and the power ratio. The EF was defined as the ratio of the average light intensities in the focal area over the average intensity in the background; the size of the focus was defined as the full-width-at-half-maximum (FWHM) value of the intensity profile across across the centre of the focus; the power ratio was defined as the fraction of total output light energy distributed in the focal area.  

Second, to compare the imaging performance of the STIN and GRIN fibres, phantoms comprising networks of carbon fibres with a nominal diameter of 7 $\mu$m were imaged over a 100 $\mu{m}$ $\times$ 100 $\mu{m}$ field-of-view with a scanning step size of 0.5 $\mu$m. 

Third, to study the dependency of imaging speed and spatial resolution on the scanning step size, PA imaging was performed With a carbon fibre network phantom. PA images on area of 50 $\mu{m}$ $\times$ 50 $\mu{m}$ were acquired with varying step sizes of 0.5, 1, 1.5, 2 and 2.5 $\mu$m, corresponding to $100^2, 50^2, 34^2, 25^2$ and $20^2$ pixels per image. To estimate the lateral resolution, an edge spread function (ESF) was first obtained by averaging across profiles at 10 adjacent positions across an edge of a carbon fibre for each PA maximum intensity projections (MIP) image in the X-Y plane, and then a line spread function (LSF) was achieved by calculating the derivative of the ESF. The lateral resolution was calculated as the FWHM value of the Gaussian fit of the LSF. Similarly, as the axial resolution is independent of the scanning step size, it was estimated by profiles across an edge of a carbon fibre in the PA maximum intensity projection image in the X-Z plane with a scanning step size of 0.5 $\mu$m.

Fourth, the capability of the probe for PAI of biological tissue was evaluated by imaging \textit{ex vivo} mouse RBCs with GRIN fibre. Mouse blood was obtained from culled mice: these procedures involving mice were ethically reviewed and carried out in accordance with the Animals (Scientific Procedures) Act 1986 (ASPA) UK Home Office regulations governing animal experimentation. To maximise the PA signal strength, the field-of-view of PA imaging was defined by the size of the acoustic focus with a side length of $\sim$50 $\mu$m. The scanning step size was set to be 1.5 $\mu$m and PA signals were averaged over 4 repeated measurements to increase the SNRs, corresponding to an imaging speed of 2.2 fps. 

Finally, the capability of dual-modal PA and fluorescence imaging was demonstrated with a phantom comprising a carbon fibre and fluorescent microspheres over an area of 100 $\mu$m $\times$ 100 $\mu$m. The carbon fibre had a nominal diameter of around 7 $\mu$m and was placed on a coverslip. A drop of a solution containing 4 $\mu$m fluorescent microspheres (TetraSpeck) was then added onto the coverslip and allowed to dry at room temperature prior to imaging. Images for the two modalities were acquired simultaneously with the same MMF (GRIN, 100 $\mu$m in diameter, 0.29 NA) and the same excitation laser. The fluorescence signals achieved from fluorescent microspheres were averaged by 4 times for higher SNRs. 


\begin{figure}
\begin{center}
\begin{tabular}{c}
\includegraphics[width=16cm]{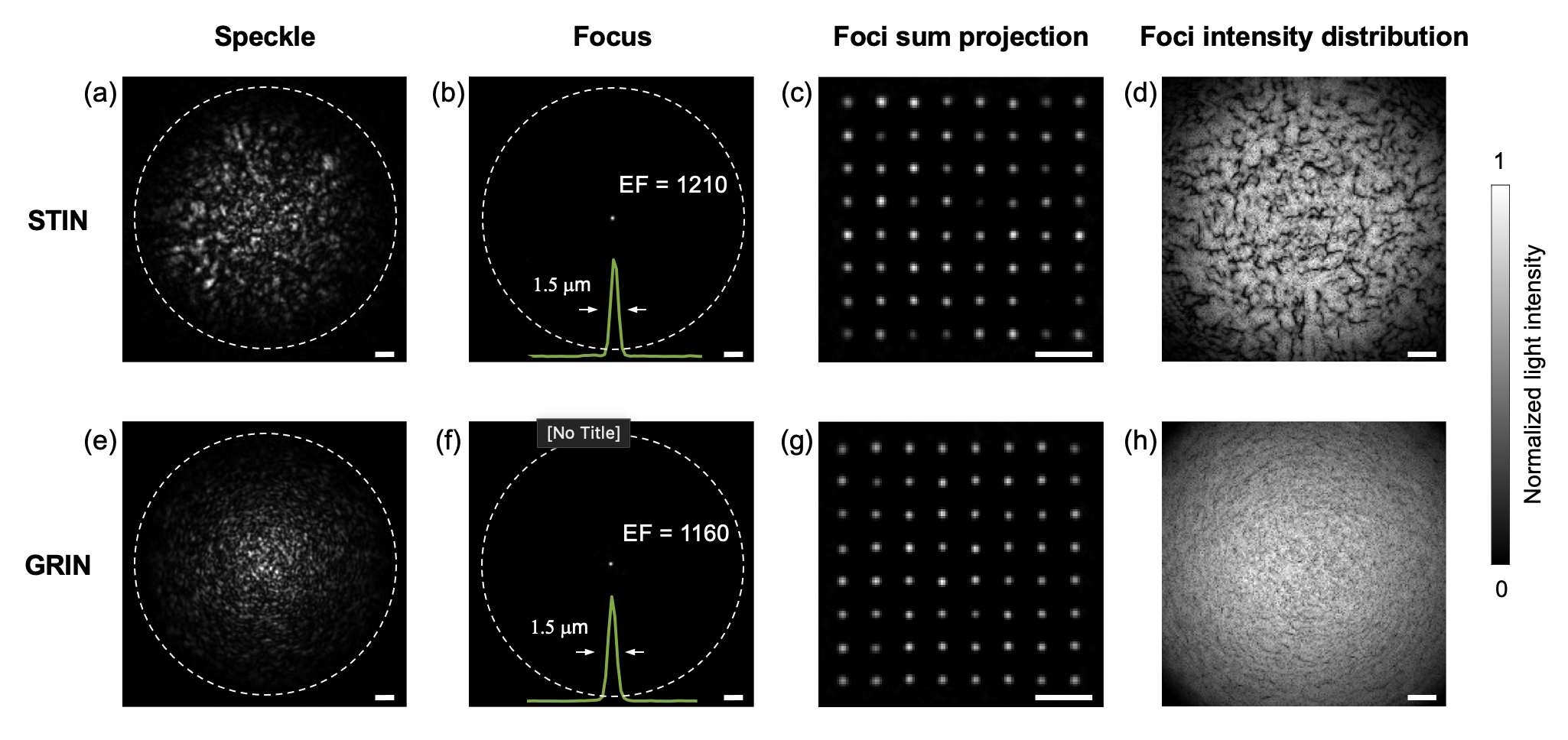}
\end{tabular}
\end{center}
\caption 
{ \label{fig:example}
Focusing performance through step-index (STIN) and gradient-index (GRIN) multimode fibres. (a-b) are images of typical output speckles with random DMD patterns as inputs, and (c-d) are images of optical foci when optimal patterns are used as inputs for STIN and GRIN fibres, respectively. The insets in (b) and (f) are lateral intensity profiles across the centres of the corresponding foci in (b) and (f). A series of foci images at different positions are summed in (c) and (g). (d) and (h) are the distribution of the peak foci intensities across the optical focal plane through the STIN and GRIN fibres, respectively.} 
\end{figure} 

\section{Results}
\subsection{Focusing performance}
Fig. 2 shows the results of light focusing performance through MMFs. When a random DMD pattern was projected to the fibres, the outputs for both fibres are random-like speckles, however, the speckles with the STIN fibre was extended to a larger region compared to those with the GRIN fibre (Fig. 2a,e). Tight optical foci were generated with the RVITM method (Fig. 2b,f) for both fibres, with an EF of 1210 for the STIN fibre and an EF of 1160 for the GRIN fibre, and the same focus size of 1.5 $\mu$m for both fibres. The power ratios for the STIN and GRIN fibres were 10.6\% and 9.1\%, respectively. Fig. 2(c,g) show the generated optical foci patterns at different spatial locations, and Fig. 2(d,h) show the peak intensity value distributions of the generated foci with a scanning step of 0.5 $\mu$m for the two fibres. It is clearly seen that the maximum values of the foci are much more uniformly distributed with the GRIN fibre compared to the STIN fibre, suggesting a better image quality for the GRIN fibre. 


\begin{figure}
\begin{center}
\begin{tabular}{c}
\includegraphics[width=15cm]{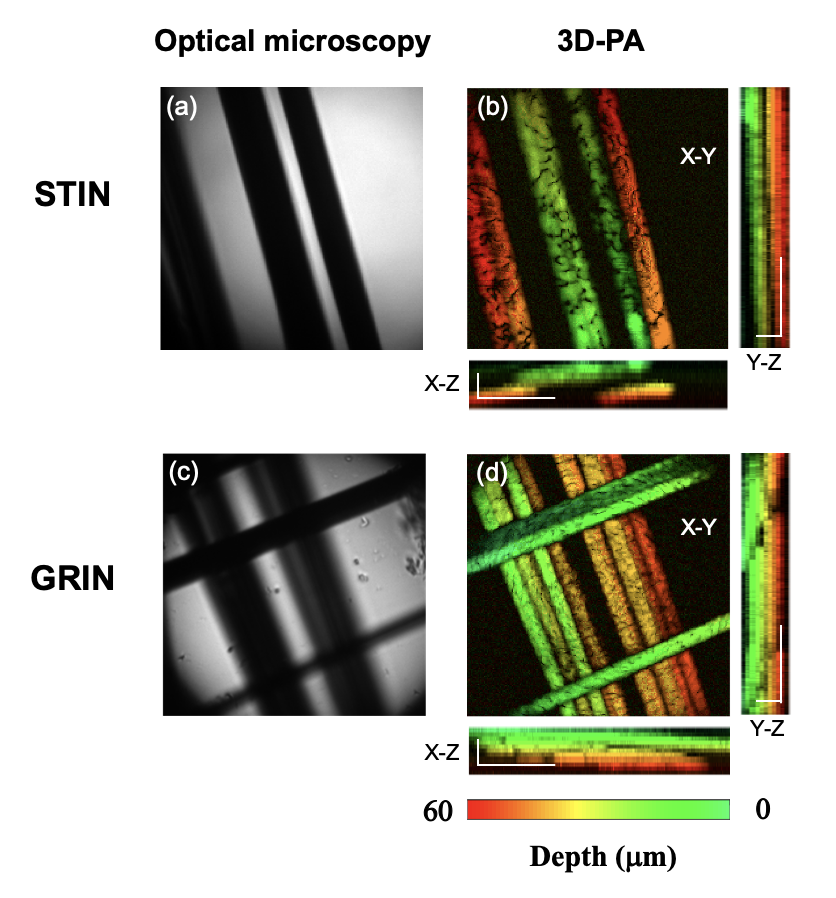}
\end{tabular}
\end{center}
\caption 
{ \label{fig:example}
Photoacoustic (PA) imaging of carbon fibres with a step-index (STIN) and a gradient-index (GRIN) multimode fibres. (b) and (d) are PA maximum intensity projection images of a carbon fibre phantom in the X-Y, X-Z and Y-Z planes with the STIN and GRIN fibres, respectively, and (a) and (c) are corresponding optical microscopy images. Depth information in the PA images are coded with false colours. Scale bars in all the X, Y and Z directions represent 30 $\mu$m.} 
\end{figure} 

\subsection{Photoacoustic imaging of carbon fibre phantoms}
Fig. 3 (a) and (c) show conventional optical microscopy images of the carbon fibre phantoms with the same field-of-view acquired in transmission mode. It can be seen that conventional optical microscopy was able to provide 2D visualisation and thus it was not able to spatially resolve individual carbon fibres in some overlapped regions. Fig. 3 (b) and (d) show the MIPs of PA images in the X-Y, Y-Z and X-Z planes for both the two MMFs. PA imaging was able to provide 3D visualisation of individual carbon fibres with high fidelity; the depth information was provided by the time-resolved PA signals from each scanning location and was coded with false colours. Compared to the STIN fibre (Fig. 3b), PA images achieved with the GRIN fibre (Fig. 3d) showed a more uniform pixel amplitude distribution across the carbon fibres regions, which was attributed to the more uniform foci intensity distribution as shown in Fig. 2 (h). 

\begin{figure}
\begin{center}
\begin{tabular}{c}
\includegraphics[width=16cm]{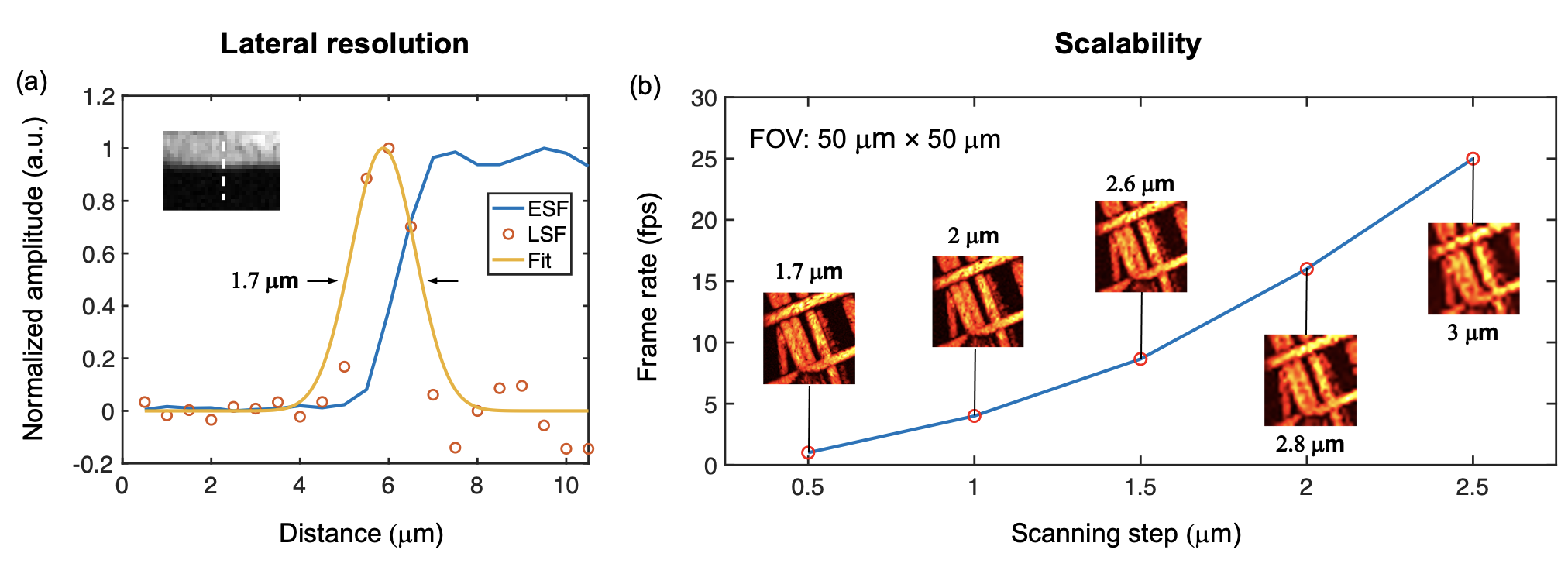}
\end{tabular}
\end{center}
\caption 
{ \label{fig:example}
Scalability of the multimode fibre-based imaging system demonstrated with imaging of a carbon fibre phantom. With an imaging field-of-view of 50 $\mu{m}$ $\times$ 50 $\mu{m}$, and the digital micromirror device running at 10kHz, the imaging speed increases with the decrease of the number of pixels per image from 1 to 25 frames per second (fps), and the resulting lateral resolution worsens from 1.7 to 3 $\mu$m.} 
\end{figure} 

Since scanning of the light focus was realised via optical wavefront shaping rather than conventional approaches with mechanical scanning mirrors, the scanning step size and field-of-review could be easily controlled by displaying corresponding DMD patterns, which enabled high scalability on the achievable imaging frame rate and spatial resolution as shown in Fig.4.  With the DMD running at 10 kHz, the frame rate of PAI varied from 1 to 25 fps with varying step sizes from 0.5 to 2.5 $\mu$m. The lateral resolution with a step size of 0.5 $\mu$m was estimated as 1.7 $\mu$m (Fig. 4a), which is consistent to size of optical focus through the same MMF (1.5 $\mu$m). With the increase of the scanning step size, the lateral resolution of sub-sampled images worsened gradually from 1.7 - 3 $\mu$m (Fig. 4b). The calculated axial resolution was $\sim$27 $\mu$m, which is consistent with the frequency bandwidth of the US transducer (27 - 63 MHz), taking into account of the frequency-dependent acoustic attenuation of water \cite{wang2014photoacoustic}. 




\begin{figure}
\begin{center}
\begin{tabular}{c}
\includegraphics[width=16cm]{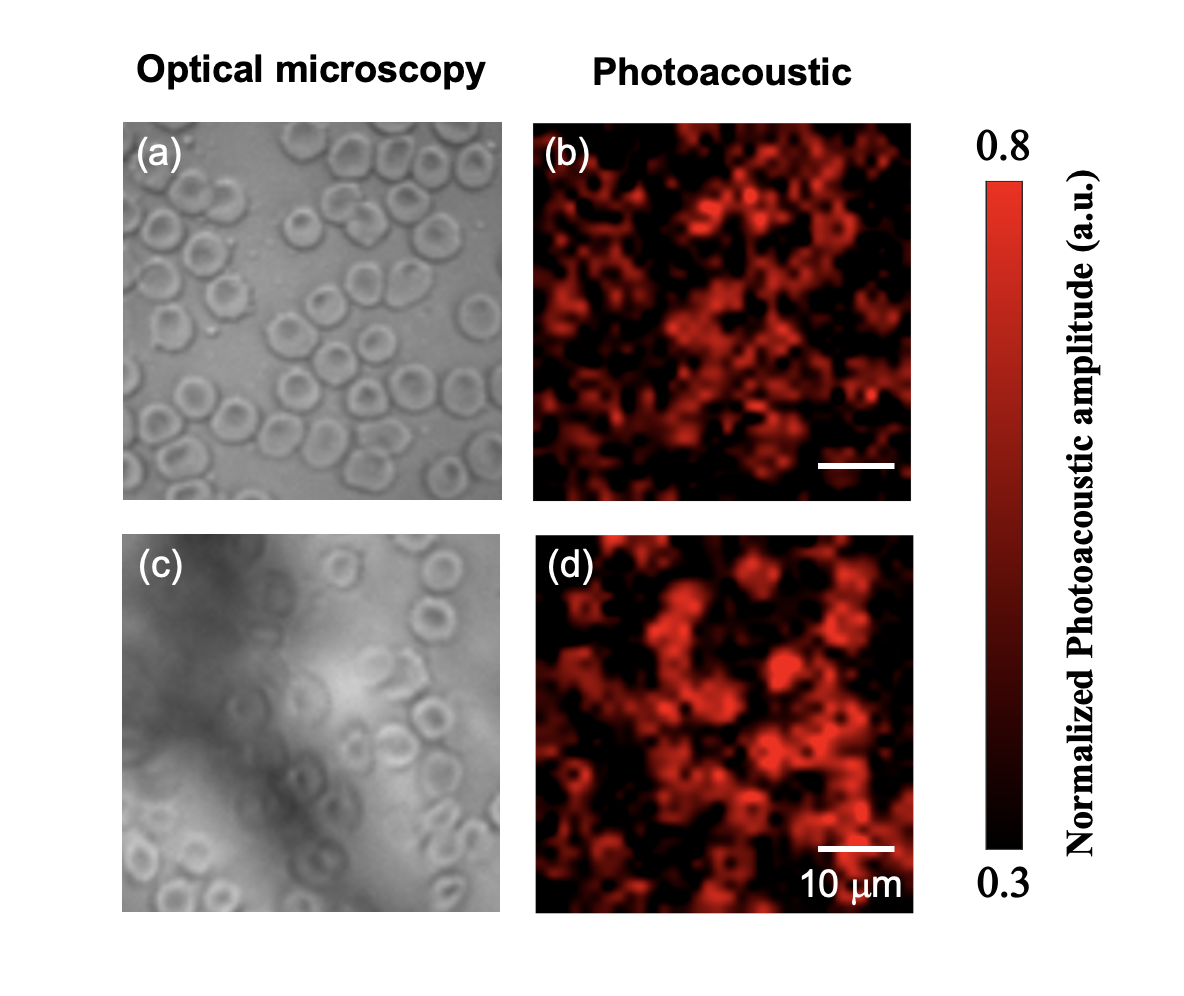}
\end{tabular}
\end{center}
\caption 
{ \label{fig:example}
Photoacoustic imaging of mouse red blood cells \textit{ex vivo}. a.u., arbitrary unit. Scale bars: 10 $\mu$m} 
\end{figure} 


\subsection{Photoacoustic imaging of mouse red blood cells}
Fig.5 (b,d) show PA MIP images in the X-Y plane achieved with averaging across 4 consecutive frames and a scanning step size of 1.5 $\mu$m. The bi-concave structures of RBCs were clearly visualised with a high consistency to the corresponding optical microscopy images (Fig. 4a,c). The image quality is comparable to those of the RBCs images obtained with benchtop PA microscopes reported in literature\cite{dong2015isometric,danielli2014label}. Each image covered an area of 51 $\mu$m $\times$ 51 $\mu$m with 34 by 34 pixels, leading to a frame rate of $\sim$2.2 fps.


\begin{figure}
\begin{center}
\begin{tabular}{c}
\includegraphics[width=15cm]{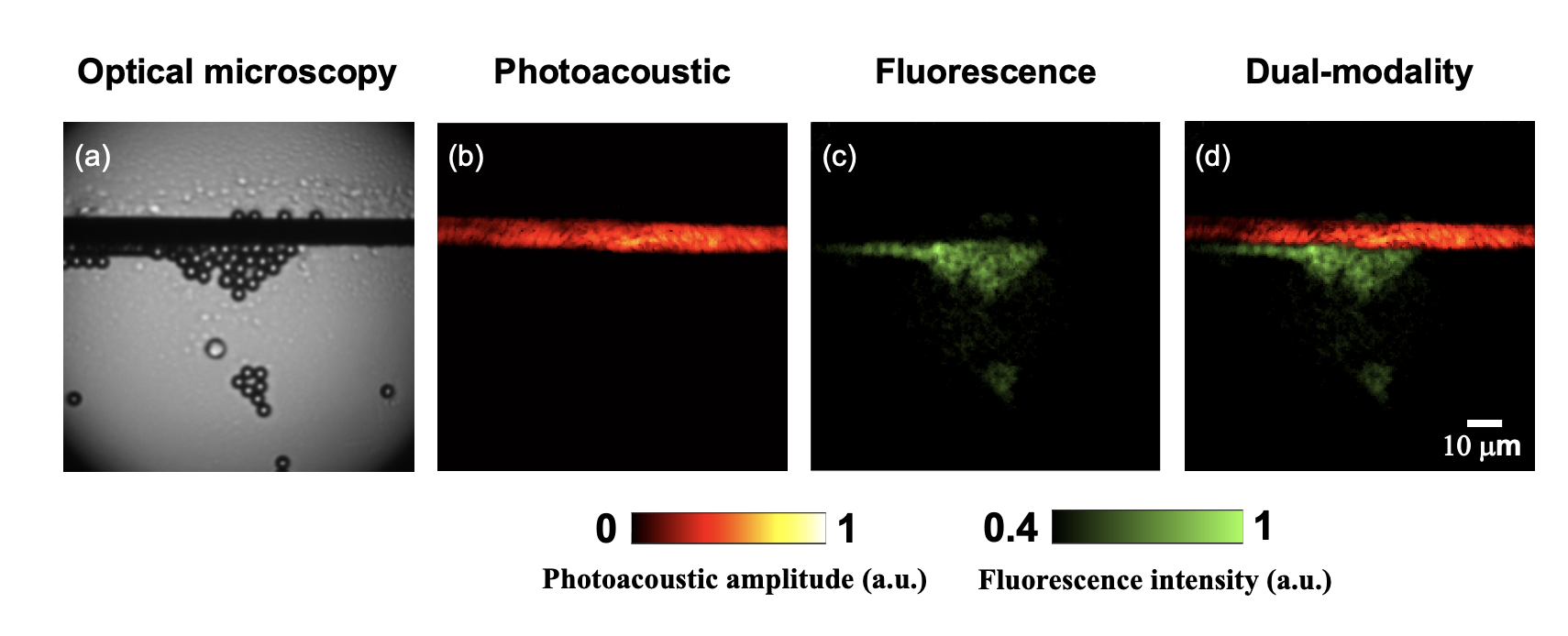}
\end{tabular}
\end{center}
\caption 
{ \label{fig:example}
Dual-modal photoacoutic and fluorescence imaging of phantoms comprising carbon fibres and fluorescent microspheres. Scale bar: 10 $\mu$m} 
\end{figure} 

\subsection{Simultaneous photoacoustic and fluorescence imaging}
As shown in Fig. 6, PA MIP images in the X-Y plane clearly visualised the carbon fibres with high fidelity and corresponded well to the corresponding optical microscopy images. On the other hand, fluorescence imaging highlighted the fluorescent microspheres as expected. However, due to the limited spatial resolution, individual fluorescent microshperes could not be resolved clearly.  
 
\section{Discussion}
In this study, we developed a high-speed dual-modal forward-viewing imaging probe that integrated simultaneous PA and fluorescence imaging capabilities at the tip of an ultrathin MMF. The performance of the imaging system in terms of spatial resolution and imaging speed was evaluated with carbon fibre phantoms, and the capacity of biological tissues imaging with PAI was demonstrated by imaging mouse RBCs with a high fidelity that is comparable to those achieved with benchtop PA microscopes \cite{dong2015isometric,danielli2014label}. Wavefront shaping with a DMD allowed raster-scanning of a focused laser beam with varying step sizes, so that the imaging speed is scalable with the scanning step size, ranging from 1 - 25 fps over an area of 50 $\mu$m $\times$ 50 $\mu$m. The capacity of high-speed imaging is attributed to the use of a DMD for light modulations. Here the frame rate of the DMD was set to be 10 kHz, corresponding to an imaging speed of scanning 10,000 image pixels per second. To our knowledge, the fastest pixel-scanning rate for MMF-based PAE in literature was 60 Hz and it took $\sim$30 s to acquire a 1800-pixel image ($\sim$0.03 fps) \cite{mezil2020single}. In comparison, our system enabled imaging over a similar field-of-view at a frame rate of $\sim$25 fps. The DMD frame rate could be further improved to 47 kHz that corresponds to an imaging frame rate of 118 fps for the same field-of-view (50 $\mu{m}$ $\times$ 50 $\mu{m}$). 


The speed of PA imaging is mainly limited by the decrease of laser pulse energy over 10 kHz. Although the RVITM method had a higher usage ratio of laser energy compared to holographic approaches \cite{Zhao21Comparison}, the laser pulse energy transmitted through the optical fibre declined to $\sim$100 nJ at a pulse repetition rate of 10 kHz. Further, the focusing enhancement obtained in this work was lower than that reported with holographic approaches with a larger number of input modes \cite{turtaev2018high} and thus resulted in lower power in light foci. Thus, signal averaging was employed for improving the SNRs of the PA signals from RBCs, which slowed down the imaging speed for biological tissue imaging. Apart from improving the laser energy, the employment of a larger number of independent micromirrors can improve the focusing enhancement, leading to a higher SNR, however, at the expense of a longer fibre characterisation time.



Compared to multi-core fibre bundle-based PA endoscopy probes \cite{hajireza2011label,ansari2018all}, the use of MMFs reduced the probe size to 140 $\mu$m in diameter and improved the lateral resolution to 1.7 $\mu$m. However, the use of MMFs also involves the challenges of fibre bending-induced system instability as changes in the fibre geometry can lead to substantial changes in the light transmission characteristics, leading to degradation of the focusing performance. Although a number of approaches have been reported to address this challenge, such as TM reconstruction based on the known fibre geometry \cite{ploschner2015seeing} and GPU accelerated real-time characterisation \cite{ploschner2014gpu}, achieving a fully flexible MMF-based imaging probe with wavefront shaping remains challenging. In this study the image quality with a GRIN fibre is found to be higher compared to that with a STIN fibre due to the more uniform distribution of the peak energies of the optical foci that is associated with the GRIN fibre. In a recent study, it was reported that GRIN fibres have a higher resistance to geometry changes on the imaging performance compared to STIN ones \cite{flaes2018robustness}. In future studies, the PA and fluorescence imaging performance will be studied with a GRIN fibre probe integrated into a medical needle and as such the geometry of the fibre could be relatively unchanged. Apart from MMF, several novel optical fibres have been investigated for higher resistance to geometry changes of the fibre. A conformationally invariant multi-core fibre with twisted cores was reported for a flexible endoscope, which allowed the light focusing with geometry changes of the fibre for digital confocal two-photon imaging \cite{tsvirkun2019flexible}. A disordered glass-air Anderson localized optical fibre was also demonstrated to be resistant to fibre bending for imaging transmission applications using deep learning \cite{zhao2018deep}.  
 
 
In this study, a conventional piezoelectric US transducer was used for US detection in a transmission mode, which restricted its use for endoscopic applications. In the future, a miniature US sensors will be attached to the MMF to receive PA signals in a reflection mode. This sensor could be a miniature piezoelectrical US sensors \cite{li2020size}, a fibre-optic Fabry-Perot sensor \cite{guggenheim2017ultrasensitive} or a micro-rings resonator US sensor \cite{westerveld2021sensitive}.

\section{Conclusions}
In summary, we developed a forward-viewing dual-modal PA and fluorescence endo-microscopy probe based on a MMF. With a fast DMD used for wavefront shaping, high-fidelity PA images of carbon fibres and mouse RBCs were acquired at high-speed for the first time. Dual-modality imaging capability was demonstrated with phantoms consisting of carbon fibre and fluorescence microspheres. With further miniaturisation of the US detector, the imaging platform has the potential to guide various minimally invasive procedures such as tumour biopsy, and nerve blocks, by providing micro-structural, functional and molecular information at high-resolution in real-time from within a medical needle.

\subsection*{Disclosures}
The authors declare that there are no conflicts of interest. T. V is co-founder and shareholder of Hypervision Surgical Ltd, London, UK. He is also a shareholder of Mauna Kea Technologies, Paris, France.

\subsection* {Acknowledgments}
The authors would like to acknowledge financial support from Wellcome Trust (203148/Z/16/Z, WT101957), and Engineering and Physical Sciences Research Council (NS/A000027/1, NS/A000049/1).



\bibliography{report}   
\bibliographystyle{spiejour}   





\end{spacing}
\end{document}